\documentclass[universe,article,accept,moreauthors,pdftex]{Definitions/mdpi} 

\firstpage{1} 
\pubvolume{7}
\issuenum{10}
\articlenumber{365}
\pubyear{2021}
\copyrightyear{2021}
\externaleditor{Academic Editor: {Norma G. Sanchez}} 
\datereceived{{31 August 2021}} 
\dateaccepted{{24 September 2021}} 
\datepublished{{29 September 2021}} 
\hreflink{https://doi.org/10.3390/\linebreak universe7100365} 

\Title{The Epoch of Reionization in Warm Dark Matter Scenarios}

\TitleCitation{The Epoch of Reionization in Warm Dark Matter Scenarios}

\Author{{Massimiliano Romanello} $^{1,2,}$*\orcidA{}, {Nicola} Menci $^{3}$\orcidB{}  and {Marco} Castellano $^{3}$\orcidC{}}

\AuthorNames{M. Romanello, N. Menci and M. Castellano}

\AuthorCitation{Romanello, M.; Menci, N.; Castellano, M.}

\address{%
$^{1}$ \quad Dipartimento
di Fisica e Astronomia, Alma Mater Studiorum Università di Bologna, Via Piero Gobetti 93/2, 40129 Bologna, Italy\\
$^{2}$ \quad Osservatorio di Astrofisica e Scienza dello Spazio di Bologna, INAF,   Via Piero Gobetti 93/3, \mbox{40129 Bologna, Italy}\\
$^{3}$ \quad Osservatorio Astronomico di Roma, INAF, Via di Frascati 33, \mbox{00040 Monte Porzio Catone, Italy}; nicola.menci@inaf.it (N.M.); marco.castellano@inaf.it (M.C.)}

\corres{Correspondence: massimilia.romanell2@unibo.it}

\abstract{In this paper we investigate how the Reionization process is affected by early galaxy formation in different cosmological scenarios. We use a semi-analytic model with suppressed initial power spectra to obtain the UV Luminosity Function in thermal Warm Dark Matter and sterile neutrino cosmologies. We retrace the ionization history of intergalactic medium with hot stellar emission only, exploiting fixed and variable photons escape fraction models ($f_{esc}$). For each cosmology, we find an upper limit to fixed $f_{esc}$, which guarantees the completion of the process at $z<6.7$. The analysis is tested with two limit hypothesis on high-z ionized hydrogen volume fraction, comparing our predictions with observational results.}


\begin{document}
\section{Introduction}
The Epoch of Reionization (EoR) marked a fundamental phase transition in the history of the Universe, during which the Intergalactic Medium (IGM) became transparent to UV photons. This phase transition is most likely caused by the energetic photons emitted by the first sources of light ever formed, i.e., galaxies and AGNs. A ``late Reionization'' scenario emerges from the combination of several probes: the Thomson scattering optical depth of cosmic microwave background (CMB) photons on the free electrons in the IGM~\cite{Planck15, Planck18,WMAP9}; the absorption due to neutral intergalactic hydrogen and the appearance of Gunn-Peterson thought in the spectra of distant QSOs~\cite{Fan2006,Wise}; the evolution of the Lyman-$\alpha$ emitters (LAEs) luminosity function~\cite{Dijkstra}; and the decreasing of the abundance of the Lyman-Break-Galaxies (LBGs) at $z>6$~\cite{Dijkstra,Pent}. Although the most recent observations indicate a late-Reionization scenario~\cite{Hoag19, Mason19, Planck18}, with the end of the EoR at $z \approx 6$, the exact contribution from different ionizing sources and the exact timeline and topology of Reionization are still unknown.

From a theoretical point-of-view, cosmic Reionization depends on non-linear and non-local phenomena, in which the physics of galaxy formation couples with the physics of gravity and radiation transport. The first process is determined by both baryonic physics and poorly known feedback effects, but also by the initial power spectrum of density fluctuations: in fact, dark matter produces the potential wells in which baryonic perturbation undergo an accelerated growth. Therefore, the study of Reionization is strongly related to the comprehension of cosmological framework in which cosmic structures form and grow. 

The currently most acknowledged cosmological model is the $\Lambda$CDM model. It is based on the contribution of the cosmological constant $\Lambda$ ($\approx$$69 \%$) and Cold Dark Matter ($\approx$$26\%$) and provides a coherent large-scale description of the Universe with respect to the available data. However, there are some possible tensions related to observations at galactic and sub-galactic scales, of the order of kpc. From N-bodies simulations an abundance of galactic sub-halo is expected, that does not match with the small number of satellites around the Milky Way~\cite{Bullock, Klypin}. Furthermore, the simulated halos density profiles are more concentrated with respect to the observed ones~\cite{Bullock}. It is possible to try to solve the so-called ``missing satellites'' problem and ``cusp-core'' problem within the $\Lambda$CDM scenario, with the aid of baryonic effects that act in the faint galaxies~\cite{Bullock}. In particular, the supernovae feedback is more efficient in the shallowest potential well, and causes the expulsion of the galactic gaseous content, while the photo-ionization feedback determines the suppression of star formation in small galaxies, because it is more difficult for their host halos to collect gas from an ionized environment~\cite{Bullock,Yue}.

The $\Lambda$CDM model postulates the existence of Dark Matter in a ``cold'' version, i.e., composed by Weakly Interacting Massive Particles (WIMPs) with $m_X>0.1$ GeV or condensates of light axions, with $m_X\approx10^{-5}$--$10^{-1}$ eV. The lack of detection of CDM candidates has suggested the possibility to investigate on alternative cosmological scenarios, based on the existence of Warm Dark Matter particles, with mass of the order of keV, which can resolve some of the problems that affect the $\Lambda$CDM model at the kpc scales. Indeed, during the structures formation, the motion and the thermal velocity of collisionless DM particles imply the deletion of cosmic perturbations which are smaller than the ``free streaming length''. While in the $\Lambda$CDM model, due to the high mass particles, all the cosmological density perturbations can become gravitationally unstable, in a WDM scenario, depending on the value of $m_X$, only perturbations above the kpc scale can collapse, producing shallower density profiles and a smaller number of low-mass halos. This, in the context of the hierarchical growth of the cosmic structures, implies a reduction in the number of faint galaxies and a delay in their formation~\cite{Menci,Dayal15}. This fact could have macroscopic consequences on the rest frame galaxy UV Luminosity Function (LF) and several studies are searching for the existence of a possible ``turn-over'' in its faint end~\cite{Yue,Castellano}.

In WDM cosmologies, the simplest approach is to consider particles that behave as ``thermal relics'', resulting from the freeze-out of DM species initially in thermal equilibrium with the early Universe. The suppression of the thermal power spectrum allows to deriving constraints to the WDM particles mass, comparing the halo mass function obtained from N-bodies simulations, with the UV luminosity function at high redshift, as performed in Corasaniti et al. (2017), who derive a lower limit $m_X>1.5$ keV~\cite{Corasaniti}. Viel et al. (2013) found $m_X>3.3$ keV, analysing the Ly$\alpha$ forests selected from a sample of QSOs~\cite{Viel}. A baryonic-independent estimation of $m_X>2.1$ keV was found by Menci et al. (2016), by using the lensing proprieties of a galaxy cluster to measure the abundances of high-z galaxies and to derive a lower limit to the dark matter halo mass function~\cite{Menci}.

A possible alternative is offered by sterile neutrinos (SN) or right-handed neutrinos, which are particles predicted in the context of Standard Model extensions. Since they are produced out-of-equilibrium, from the oscillations of active neutrinos, they are characterized by a non-thermal power spectrum, which depends both on mass and on $\sin(2\theta)$, where $\theta$ is the mixing angle~\cite{Merle}. Sterile neutrinos can constitute an example of radiatively decay dark matter, potentially explaining the discovery of a 3.5 keV emission line observed towards the centre of galaxy and in Perseus, Coma and Ophiuchus Clusters~\cite{Iakubovskyi}.

Some recent works have treated galaxy formation and evolution in a WDM framework. Dayal et al. (2015) found that the suppression of small-scale structures leads to a delayed and more rapid stellar assembly in thermal $1.5$ keV WDM~\cite{Dayal15}. This result was confirmed also for sterile neutrino cosmologies by Menci et al. (2018), which indicates that models with suppressed power spectra are characterized by a delay in the stellar mass growth history, ranging from 500 Myr to 1 Gyr~\cite{Menci18}. Reionization in different cosmologies was already discussed in Dayal et al. (2017). Their study pointed out that, in CDM models, the bulk of the ionizing photons is produced by systems with $-15<M_{UV}<-10$ and $M_{halo}\lesssim10^9M_\odot$, while in WDM case they registered a shift towards $M_{halo}\gtrsim10^9M_\odot$ and $-17<M_{UV}<-13$, due to the effect of suppression. However, the faster WDM galaxy assembly implies that the reheating of IGM is almost finished at $z\approx5.5$--$6$~\cite{Dayal17}. The paper by Carucci et al. (2018) adopted N-body simulations to produce the high-redshift halo mass function. They used an abundance matching method to convert halo masses into UV galaxy magnitudes, between $6<z<10$. Then, they have assumed galaxies as the main source of ionizing photons, finding a very similar reionization history for CDM, WDM, late-forming dark matter and ultralight axion dark matter models~\cite{Carucci}.

In this paper we investigate how Reionization scenarios are affected by early galaxy formation in WDM cosmologies. We have used the theoretical model by Menci et al. (2018)~\cite{Menci18}, where the collapse history of dark matter halos is modelled through the Extended Press-Schechter (EPS) formalism and baryonic processes taking place in each halo are included through physically motivated analytical recipes. We have focused on three thermal WDM scenarios, with masses of $2$--$3$--$4$ keV and five sterile neutrino models, with a 7 keV mass, but characterized by different lepton asymmetry parameter, $L_6$, strongly related to the mixing angle~\cite{Lovell16}.

The paper is organized as follows. Section \ref{sec_Materials_and_method} is dedicated to the semi-analytic model and to our hypotheses on Reionization history: Section \ref{subsez_soppressione_power_spectrum} provides a description of the suppression function of the original CDM power spectrum, in Section \ref{subsec_modelling_reioniz} we outline how to reconstruct the evolution of hydrogen filling fraction with different models of escape fraction. In Section \ref{sez_results} we describe the properties of our ionizing sources: in \mbox{Section \ref{subsec_funzione_lum}} we show the UV Luminosity function in different WDM scenarios, in Section \ref{sez_reconstructiong_reionization_history} we evaluate the contribution of faint and bright galaxies with a fixed escape fraction model. Section \ref{sec_discussion} is devoted to results and discussion, which includes the comparison with some authors in literature and with selected observational data (Sections \ref{sez_vincoli_osservativi} and \ref{sez_confronto_finale}), the description of Reionization process with galaxies (Section \ref{sec_reion_gal_only}). Finally, in Section \ref{sec_conclusioni} we have the conclusions. In this work we adopt the following cosmological parameters: \mbox{$H_{0}=70$ km/s/Mpc}, $\Omega_\Lambda=0.7$, $\Omega_m=0.3$ and $\Omega_b=0.045$.
\section{Materials and Methods}
\label{sec_Materials_and_method}
\subsection{Semi-Analytic Model}
\label{subsez_soppressione_power_spectrum}
In our study we use the semi-analytic model developed by Menci et al. (2018), to which we refer for further informations~\cite{Menci18}. The model retraces the collapse of dark matter halos through a Monte Carlo procedure on the basis of the merging history given by EPS formalism, at $0<z<10$~\cite{Menci18}. In this framework, the DM structures formation is determined by the power spectrum: the WDM $P(k)$ is computed by the suppression of the CDM one,  due to the particles free streaming at kpc scale. A half-mode wavenumber is defined, as the $k_{hm}$ at which the transfer function $T_{WDM}(k)$ is equal to $1/2$~\cite{Bode,Schneider,Viel2005}. For thermal WDM, the power spectra ratio is related to the WDM particle mass $m_X$:
\begin{equation}
    T_{WDM}(k)=\left[\frac{P_{WDM}}{P_{CDM}}\right]^{1/2}=[1+(\epsilon k)^{2 \mu}]^{-5/\mu},
\label{eq_funz_trasferimento}
\end{equation}
with $\epsilon$ equal to:
\begin{equation}
    \epsilon=0.049\left[\frac{\Omega_X}{0.25}\right]^{0.11}\left[\frac{m_X}{keV}\right]^{-1.11}\left[\frac{h}{0.7}\right]^{1.22}Mpc\: h^{-1},
\end{equation}
where $\Omega_X$ is the density parameter of DM and $\mu=1.12$~\cite{Menci,Menci18}. From Equation (\ref{eq_funz_trasferimento}) we can also define a half-mode mass $M_{hm}$:
\begin{equation}
\label{halfmodemass}
    M_{hm}=\frac{4\pi}{3}\rho_m \left [\pi\epsilon(2^{\mu/5}-1)^{-1/2\mu}\right]^3.
\end{equation}

Conversely, for sterile neutrino based cosmological scenarios, we refer to $M_{hm}$ from Lovell et al. (2020), obtained comparing CDM and WDM simulations performed within the same cosmic volume and in which the parameterization of the WDM halo mass function is given by $R_{fit}$~\cite{Lovell}:
\begin{equation}
\label{fitting}
    R_{fit}=\frac{n_{WDM}}{n_{CDM}}=\left(1+\left(\alpha\frac{M_{hm}}{M_{halo}}\right)^\beta\right)^\gamma,
\end{equation}
where $n_{CDM}$ and $n_{WDM}$ are the differential halo mass functions and $M_{halo}$ is the halo-mass. The numerical value of $\alpha$, $\beta$ and $\gamma$ coefficients changes if we consider central ($\alpha=2.3$, $\beta=0.8$, $\gamma=-1.0$) or satellite halos ($\alpha=4.2$, $\beta=2.5$, $\gamma=-0.2$)~\cite{Lovell}. 

We perform our analysis with five different sterile neutrino models, with a mass of $7.0$ keV, labelled according to the lepton asymmetry number ($L_6$), which is indicated in the last part of the name. For example, $L_6= 120$ is named LA120, $L_6= 8$ is named LA8 and so on. Among them, the models LA9, LA10 and LA11 are based on decaying-particles that are compatible with the X-ray $3.55$ keV emission line observed in galaxy clusters~\cite{Lovell}. In \mbox{Table \ref{tabellamodelli}} we also present $M_{hm}$ of three thermal WDM scenarios, with $m_X=$ 2--3--4 keV, computed using Equation (\ref{halfmodemass}).

\begin{specialtable}[H] 
\caption{Half-mode mass ($M_{hm}$) for each Dark Matter model analysed in this work.}
\label{tabellamodelli}
\setlength{\cellWidtha}{\columnwidth/2-2\tabcolsep+0.0in}
\setlength{\cellWidthb}{\columnwidth/2-2\tabcolsep+0.0in}
\scalebox{1}[1]{\begin{tabularx}{\columnwidth}{
>{\PreserveBackslash\centering}m{\cellWidtha}
>{\PreserveBackslash\centering}m{\cellWidthb}}
\toprule
\textbf{DM Model}	& \textbf{\boldmath{$M_{hm}$} (\boldmath{$M_\odot$})}	\\
\midrule
    LA8   & $1.3\times10^8$  \\
    LA9   & $2.6\times10^8$  \\
    LA10  & $5.3\times10^8$  \\
    LA11  & $9.2\times10^8$  \\
    LA120 & $3.1\times10^9$  \\
    WDM 2 & $1.6\times10^9$ \\
    WDM 3 & $4.1\times 10^8$ \\
    WDM 4 & $1.6\times 10^8$ \\
\bottomrule
\end{tabularx}}
\end{specialtable}
\subsection{Modelling Reionization}
\label{subsec_modelling_reioniz}
In the last few years, the completion of deep surveys with efficient IR telescopes has enabled the study of number density and emission properties of distant galaxy populations. In particular, gravitational lensing turned out as an effective tool to achieve the detection of high-redshift galaxies, leading to robust constraints on the observable UV luminosity function at both the bright and the faint end. The UV LF is particularly important because it is strongly related to star formation rate~\cite{BouwensLF,Finkelstein15} and so on the existence of a hot stellar population that acts as a ionizing source responsible of Reionization. The progressive steepening of the UV LF faint-end slope at high redshift confirms the relevant role of faint galaxies in the reheating of IGM~\cite{Atek,Livermore}. 

The semi-analytic model associates a galactic luminosity to each halo, depending on cooling process and merging history. The gas in the halo, initially set to have a density given by the universal baryon fraction and to
be at the virial temperature, cools due to atomic processes and settles into a rotationally supported disk. Then, the cooled gas is gradually converted
into stars, with a SFR given by: $\dot{M}_* = \frac{M_{gas}}{\tau_*}$, according to the Schmidt-Kennicut law with a gas conversion time scale $\tau_* = q \tau_d$, proportional to the dynamical time scale $\tau_d$ through the free parameter $q$~\cite{Menci18}. Moreover, galaxy interactions occurring in the same host halo may induce the sudden conversion of a fraction $f$ of cold gas into stars on a short time-scale given by the duration of the interaction~\cite{Menci18}. Feedback phenomena due to supernovae, AGNs and photoionization are also included, as described by Menci et al. (2018)~\cite{Menci18}. Finally, the luminosity produced by the stellar populations is computed by assuming a Salpeter IMF~\cite{Menci18}. Thus, there is no universal relationship between $M_{UV}$ and $M_{halo}$, and from the suppressed halo mass function we can calculate the galaxy luminosity function in any cosmological scenario. In our analysis, we integrate the rest-frame UV ($\sim$1400  \AA) 
dust-corrected LF between the limits $M^{lim}_{UV}= [-25, -12]$, in order to obtain the corresponding luminosity density:
\begin{equation}
\label{densitàluminosità}
    \rho_{UV}=\int^{M^{lim}_{UV}}{\mathrm{d}M_{UV}\frac{\mathrm{d}N} {\mathrm{d}M_{UV}}R_{fit}L_{UV}},
\end{equation}
which is dominated by the contribution of systems with $M_{UV}\geq -20$ (see Section \ref{sez_reconstructiong_reionization_history}). Absolute magnitude is linked to UV luminosity through: $M_{1400}=51.53-2.5 \log (L_{1400}/$ $(erg\: s^{-1} Hz^{-1}))$~\cite{Lapi,Kuhlen}.

The number density of UV photons that actively participate to hydrogen ionization process is obtained by multiplying for two quantities~\cite{Finkelstein}: 
\begin{equation}
    \dot{N}_{ion}=f_{esc} \xi_{ion} \rho_{UV}.
\end{equation}

The ionizing photon production efficiency ($\xi_{ion}$) is expressed in $Hz/erg$ units and it describes how efficiently is possible to get UV ionizing photons from an UV continuum radiation field. This quantity depends on different astrophysical parameters which potentially could alter the temperature of emitting stellar population, as metallicity, initial mass function (IMF), age and binarity~\cite{Finkelstein}. During our analysis, we explore some fixed value around $\xi_{ion}\approx10^{25} Hz/erg$ (see Section \ref{sez_implications_fesc}).

Finally, the escape fraction $f_{esc}$ converts the intrinsic ionizing emissivity $\dot{N}_{ion, intrinsic}=\xi_{ion} \rho_{UV}$ into an effective one. It is defined as the fraction of ionizing photons that can escape from the source galaxy instead of being reabsorbed inside it and which therefore actively participates in the ionization of the IGM. Due to the difficulty in detection of Ly-C photons ($\lambda<912 $ \AA) for $z>4$ and to the scarce knowledge of the interstellar medium (ISM) geometry of high-z galaxies, this parameter is not actually well constrained and it summarizes most of our uncertainties about the EoR.

In our study, we model the Reionization history with different values of $f_{esc}$. Fixed escape fraction is useful to broadly characterize the Reionization history, although a universal value for $f_{esc}$ is highly unrealistic. Galaxies represent in fact a very complex sample, with large differences in morphology, mass, luminosity, colour, SFR, gas and dust content, age and metallicity. Nevertheless, the investigation of the degenerate quantities $f_{esc}\xi_{ion}$, which drive the Reionization process, can yield to interesting upper limits to the \mbox{escape fraction}.

Once obtained $\dot{N}_{ion}$, the equation that accounts for ionization and recombination, which regulates the evolution of the hydrogen filling fraction $Q_{HII}$ is: 
\begin{equation}
\label{eq_fillingfrac}
    \dot{Q}_{HII}=\frac{\dot{N}_{ion}}{\bar{n}_H}-\frac{Q_{HII}}{t_{rec}},
\end{equation}
where the comoving hydrogen mean density is computed as $\bar{n}_H\approx2\times10^{-7}(\Omega_b h^2/0.022)$ cm$^{-3}$ and the recombination time-scale is $t_{rec}\approx3.2$ Gyr $[(1+z)/7]^{-3}C^{-1}_{HII}$~\cite{Lapi}. We consider case B of recombination, in which electrons fallen to the ground level generate ionizing photons that are
re-absorbed by the optically thick IGM, having no consequences on the overall ionization balance~\cite{Wise}. We treat the evolution of the clumping factor $C_{HII}$ with redshift, due to the effect of UVB generated by Reionization, according to~\cite{Haardt, Pawlik}:
\begin{equation}
\label{eq_clump}
    C_{HII}=1+43z^{-1.71}.
\end{equation}

In Section \ref{sect_clumping_factor} we make a comparison with fixed $C_{HII}=3$ and with a $C_{HII}^{DM}$ model dependent on cosmology.

After the reconstruction of the Reionization history, we use the redshift evolution of the filling fraction to compute the integral:
\begin{equation}
\label{eq_taues}
    \tau_{es}(z)=c\sigma_T\bar{n}_H\int_0^z Q_{HII}(z')(1+z')^2 \left(1+\frac{\eta Y}{4X}\right)H^{-1}(z')\mathrm{d}z',
\end{equation}
in which helium is singly-ionized ($\eta=1$) at $z>4$ and doubly-ionized ($\eta=2$) at \mbox{$z<4$}~\cite{Bouwens15}. Then, the electron scattering optical depth has been compared with observational constraints on $\tau_{es}$ obtained, from CMB anisotropy, by Planck and WMAP.
\section{Properties of Ionizing Sources}
\label{sez_results}
\subsection{UV Luminosity Function}
\label{subsec_funzione_lum}

\begin{figure}[H]
 \hspace{-10pt} \includegraphics[width=13.5 cm]{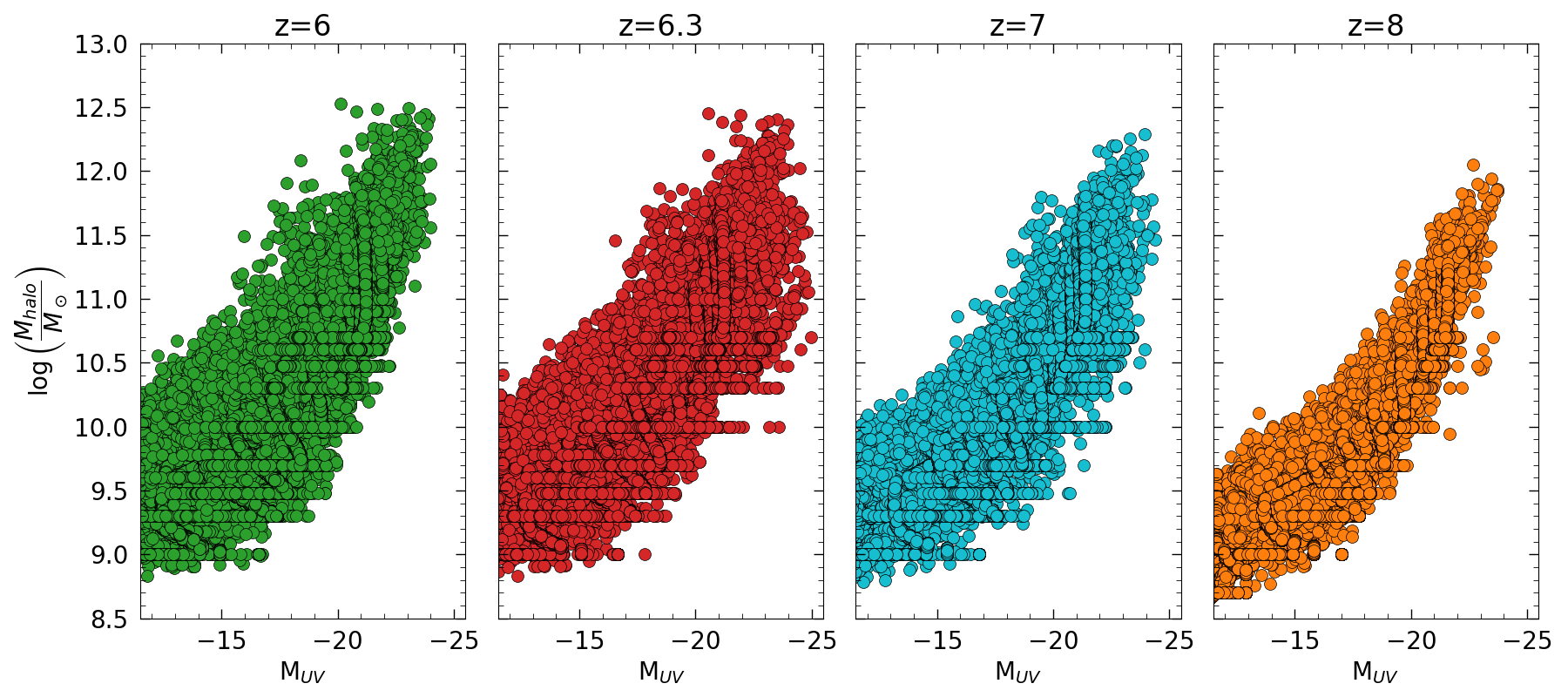}
\caption{$M_{halo}-M_{UV}$ relationship for dust attenuation-corrected stellar populations, at $6<z<8$.
\label{fig_magnitudine_massa_aloni}}
\end{figure}

Figure \ref{fig_magnitudine_massa_aloni} describes the $M_{halo}-M_{UV}$ correlation for dust-corrected UV $1400$ \AA~emission, derived from the semi-analytic model, at $6<z<8$. The UV magnitude is progressively shifted with time towards lower values: galaxies become more luminous and their overall ionizing photons budget increases. A similar tendency affects $M_{halo}$, because of merging between halos. We note also a widening in the $M_{halo}-M_{UV}$ relationship, due to the appearance of new systems and to their different physical evolution.

\begin{figure}[ht]
\widefigure
\includegraphics[width=16.5 cm]{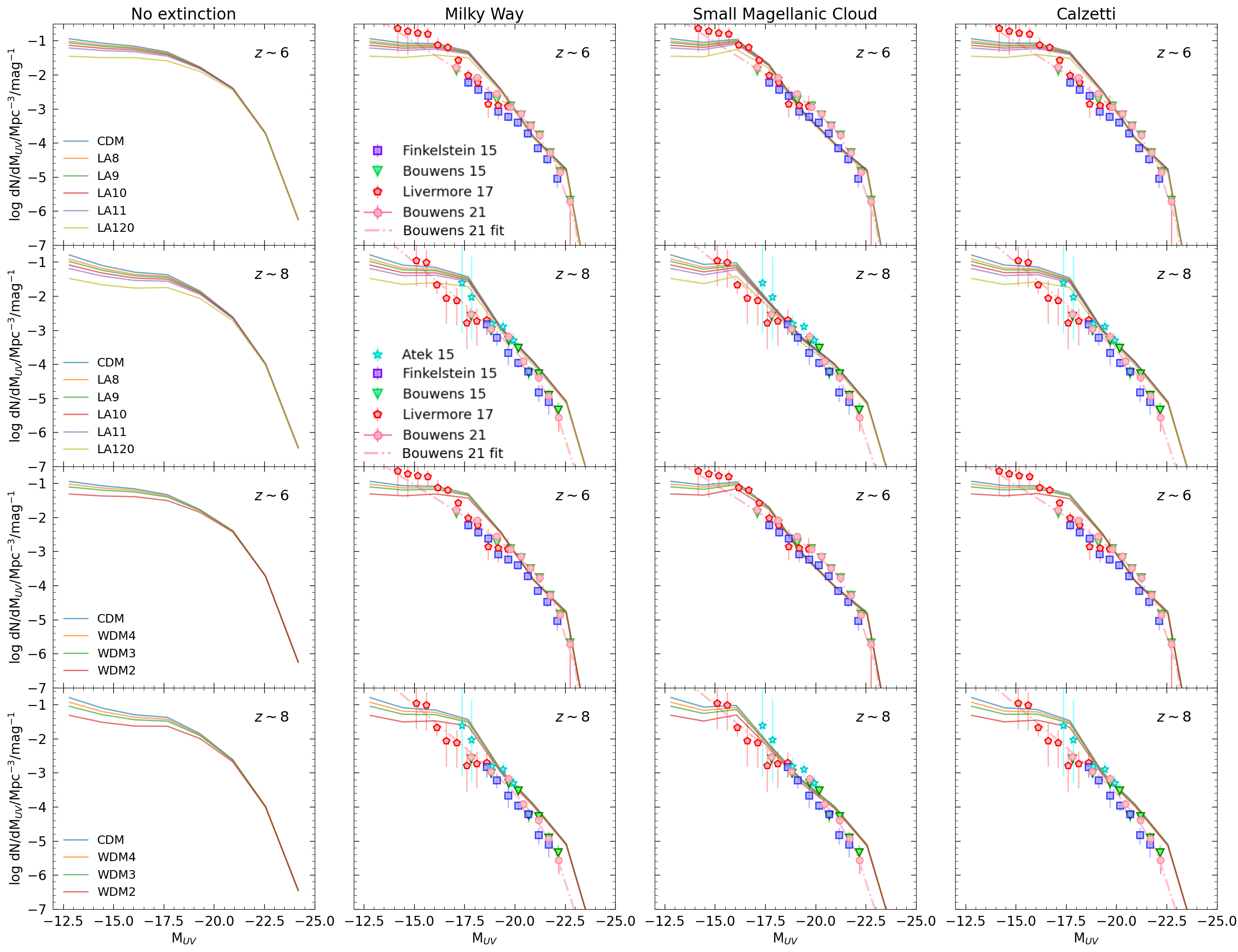}
\caption{UV LF in different cosmologies, computed using the semi-analytic model with different dust extinction laws and comparison between CDM, 2--3--4 keV thermal WDM and sterile neutrinos. Observational constraints are taken from Atek et al. (2015)~\cite{Atek} (cyan stars), Livermore et al. (2017)~\cite{Livermore} (red pentagons), Bouwens et al. (2015)~\cite{BouwensLF} (green triangles), Bouwens et al. (2021)~\cite{Bouwens21} (pink circles) and Finkelstein et al. (2015)~\cite{Finkelstein15} (blue squares). The pink dot-dashed line represents the Schechter fit derived in {Bouwens et al. (2021)}~\cite{Bouwens21}. 
\label{fig_funzioneluminosità}}
\end{figure}

In Figure \ref{fig_funzioneluminosità} we show the UV rest frame luminosity function at $1400$ \AA~in different cosmological scenarios. It is computed with the semi-analytic model within the magnitude interval $-25 < M_{UV} < -12$. Milky
Way, Small Magellanic Cloud, Calzetti extinction curves are included. 
Here, the dust extinction is treated assuming the dust optical depth to be proportional to the metallicity of the cold gas and to the disk surface density. We refer to Menci et al. (2002) for further details~\cite{Menci2002}.

From a physical point of view, the free-streaming of dark matter particles suppresses, on average, the formation of the smallest halos, whereas the most massive perturbations are still able to collapse. For this reason, comparing CDM with sterile neutrino and thermal WDM cosmologies, we find a reduction in the comoving galaxy density at higher absolute UV magnitude, with the increasing of the half-mode mass, $M_{hm}$. Although this effect is more evident for thermal WDM with $m_X=2$ keV (WDM2) and for sterile neutrino LA120, it is still insufficient to discriminate between models, with the current observational constraints on UV LF. 

WDM galaxy formation is also characterized by an initial delay in the halos collapse and consequently in the star formation process, which manifests itself at lower redshift~\cite{Calura}. Star formation is however more rapid with respect to CDM scenario~\cite{Dayal15}; this fact progressively reduces the gap between cosmologies with the age of universe, as we can see from Figure \ref{fig_funzioneluminosità}, looking at $z=8$ and $z=6$ LFs. The result is that differences in the UV LFs become negligible in the local universe.

The various UV LFs with dust extinction allow a first comparison with some recent high-z observations performed by Hubble Space Telescope. First, the paper by \mbox{Atek et al.} (2015) is based on the lensing properties of three galaxy cluster: A2744, MACS 0416 and MACS 0717, and on a sample of 25 galaxies at $z \approx 8$~\cite{Atek}. Finkelstein et al. (2015) indeed analyze the rest frame  UV LF between $4 < z < 8$, with a set of approximately 7500 candidates, obtained by exploiting the galaxy clusters Abell 2744 and\mbox{ MACS J0416.1-2403~\cite{Finkelstein15}.}

\clearpage
\end{paracol}
\nointerlineskip
\begin{paracol}{2}
\switchcolumn

We also report the paper by Livermore et al. (2017), in which a catalogue of \mbox{167 objects} is used and by Bouwens et al. (2015), based on observations of 10,000 galaxies at $z \geq 4$, whose determinations were updated and refined in Bouwens et al. (2021)~\cite{BouwensLF,Bouwens21,Livermore}. These studies have found a steepening in the faint-end of the UV LF, which enforces the role of faint galaxies in the Reionization process. However, the evidence of a turn-over in the luminosity function, which could be explained in WDM cosmologies, is still not detected~\cite{Yue,Castellano}.

In Figure \ref{fig_funzioneluminosità} we can see that, despite the still large uncertainties in the data, the Small Magellanic Cloud extinction law exhibits a better agreement between $-19<M_{UV}<-16$. We should remember, however, that the evolution of the filling fraction is computed from the dust attenuation-corrected ionizing emissivity, since all the effects that contribute to reduce the number density of ionizing photons are summarized into $f_{esc}$.

Starting from the $M_{1500-1600}$ \AA~LFs available in literature, we apply an appropriate correction in order to obtain $M_{UV}$ with an effective rest-frame wavelength of $1400$ \AA. The correction is based on the UV continuum slope $\beta$ ($f_\lambda\approx \lambda^\beta$), measured by Bouwens et al. (2014) and depends on absolute magnitude and redshift~\cite{Bouwens14}. The current measurements and uncertainties do not allow us to exclude any cosmological scenario, neither the most suppressed ones. 
\subsection{The Role of Bright and Faint Galaxies in the Epoch of Reionization}
\label{sez_reconstructiong_reionization_history}

In Figure \ref{fig_photonsfraction} we plot the integrated ionizing photons ratio:
\begin{equation}
r_{phot}(<M_{lim}^{UV})=\dfrac{\dot{N}_{ion}(M_{UV}<M_{UV}^{lim})}{\dot{N}_{ion,tot}}
\end{equation}
in which we compute $\dot{N}_{ion,tot}$, using Equation (\ref{densitàluminosità}) between intrinsic $M_{UV}^{sup}=-12$ and $M_{UV}^{inf}=-25$, while the numerator is obtained by varying the upper limit of the integral from $-24$ to $-12$, including so the photons from progressively dimmer sources, until the unity is reached. 

\begin{figure}[H]
\includegraphics[width=13.4 cm]{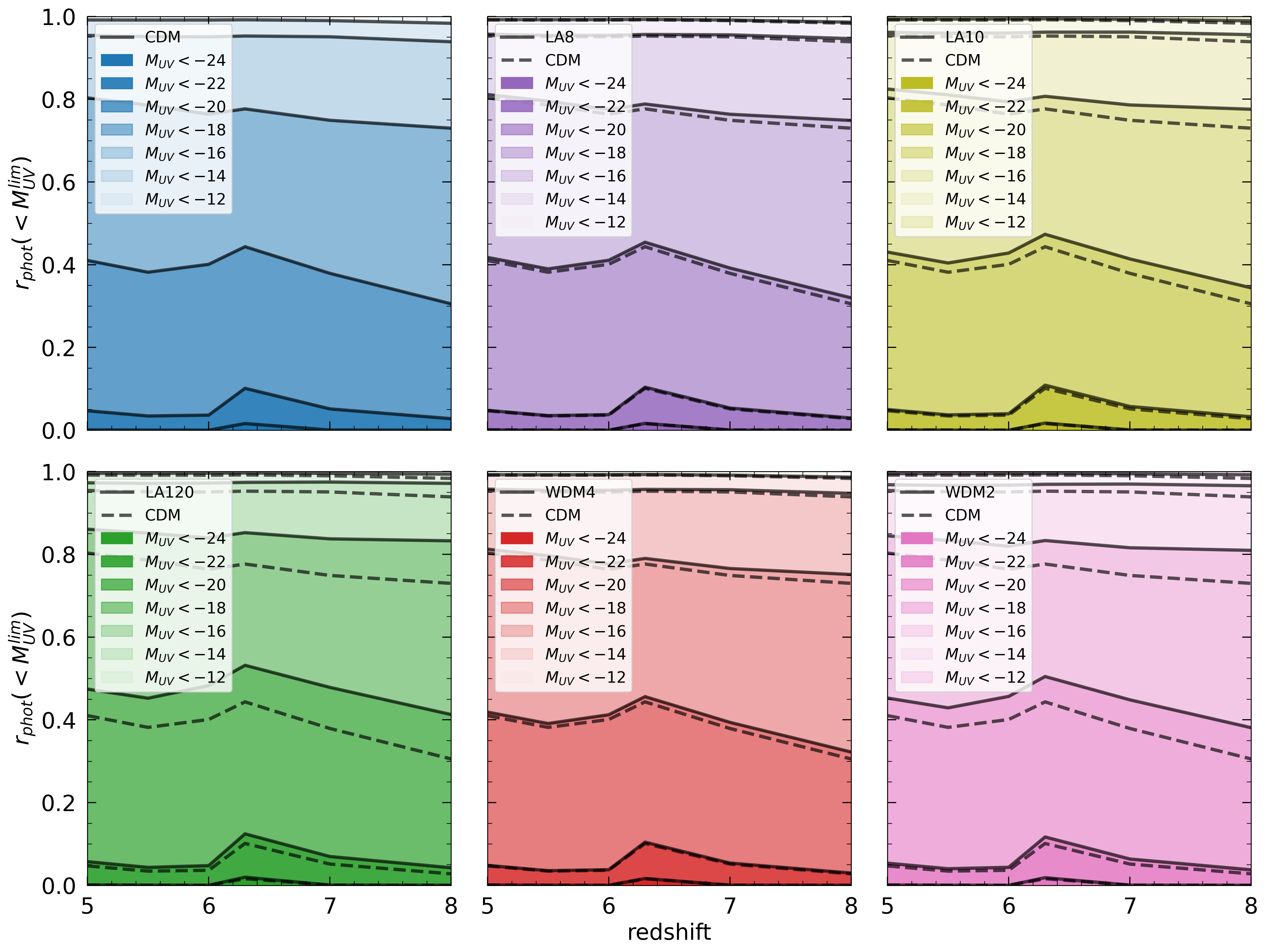}
\caption{The multiple panels show the integrated photons ratio $r_{phot}(<M_{lim}^{UV})$, where $\dot{N}_{ion,tot}$ is computed by integrating Equation (\ref{densitàluminosità}) between intrinsic $M_{UV}^{sup}=-12$ and $M_{UV}^{inf}=-25$. We compare with CDM two thermal WDM cosmologies (WDM3 is an intermediate case between WDM4 and WDM2), and three sterile neutrino cosmologies (here LA10 is the only representative scenario for radiatively decay Dark Matter, which is compatible with the 3.5 keV emission line observed in galaxy clusters).} \label{fig_photonsfraction}
\end{figure}

From Figure \ref{fig_photonsfraction} we can identify two important features, through which we can understand the role and the different contribution of faint and bright galaxies during EoR.

The first is the increasing of the relative contribution of the brightest systems (respectively with intrinsic $M_{UV}^{sup}<-24$, $M_{UV}^{sup}<-22$ and $M_{UV}^{sup}<-20$) with the age of universe. In the $\Lambda$CDM model, $r_{phot}(<-22)$ passes from $2.8\%$ at $z=8$, to $10\%$ at $z=6.3$. In parallel, for $M_{UV}^{sup}=-20$ we have a raise from $31\%$ at $z=8$ to $44\%$ at $z=6.3$. We can interpret this trend in the light of the hierarchical growth of cosmic structures: merging phenomena between galaxies give origin to more massive and brighter structures, increasing their overall contribution. However, the role of faint galaxies in the Reionization process is \mbox{still predominant.} 

The second issue to be highlighted derives from a comparison between different cosmological scenarios, which reveals that WDM models present a relative $\dot{N}_{ion}$ higher than the CDM ones. Again, the reason resides in the effect of free-streaming, which determines a suppression in the number density of the faint-galaxies and so a decreasing in their relative contribution for each $M_{UV}^{sup}$. As found by Dayal et al. (2017), the progressive suppression of low-mass halos in WDM models produces a shift in the Reionization population to larger halo and galaxy masses~\cite{Dayal17}.

The difference between cosmologies is summarized in the half-mode mass and is not negligible: if we compare CDM with LA8 and WDM4, at $z=8$ it values~$\approx$~1--2$\%$ , respectively for $M_{UV}^{sup}=-20$ and $-18$, but it increases to $8-10\%$ for WDM2 and LA120. Finally, we noted that the continue (WDM) and the dashed (CDM) lines in Figure \ref{fig_photonsfraction} approach each other with time; for example, at $z=5$ the differences between CDM and WDM2-LA120 reduce respectively to 4--6$\%$. Again, we can interpret this result by looking at the evolution of the UV LFs with $z$.

The same physics explains also the evolution over time of:
\begin{equation}
r_{phot}(<M_{lim}^{halo})=\dfrac{\dot{N}_{ion}(M_{halo}<M_{halo}^{lim})}{\dot{N}_{ion,tot}}.
\end{equation}

In Figure \ref{fig_photonsfraction_halo}, the particles free-streaming determines the reduction of the low-mass halos contribution in WDM cosmologies, while in CDM case, the importance of the most massive sources (with $M_{halo}>10^{10.5} M_\odot$) increases from $22\%$ at $z=8$ up to $48\%$ at $z=6$, due to merging phenomena. Again, the gap between cosmologies depends on the suppression of the power spectrum, i.e., on the half-mode mass; for example, systems with $M_{halo}<10^{9.5}M_\odot$, at $z=8$, contribute up to $12\%$ in CDM and up to $6\%$ in LA120.

\begin{figure}[H]
\includegraphics[width=13.4 cm]{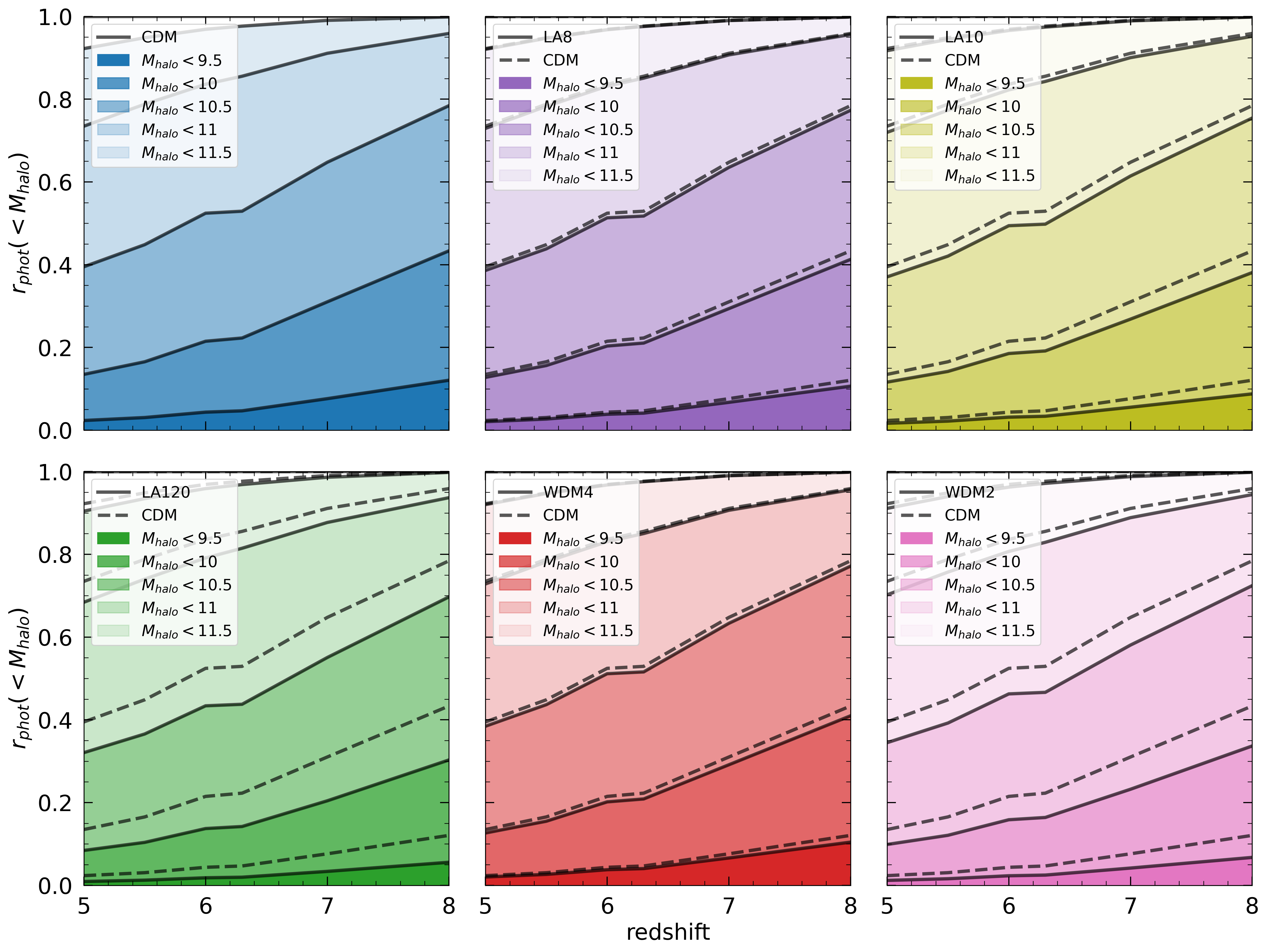}
\caption{Same as Figure \ref{fig_photonsfraction}, but with the integrated photons ratio $r_{phot}(<M_{lim}^{halo})$.} \label{fig_photonsfraction_halo}
\end{figure}

The current analysis is based on intrinsic UV luminosity and it is independent from the dust extinction, which is summarized in the escape fraction value: in fact $f_{esc}$ appears only as a multiplicative constant, so it simplifies in the ratio between $\dot{N}_{ion}$. Conversely, if we consider other escape fraction dependencies (see Section \ref{sec_implications_reionization}), we could expect a more various behaviour.
\section{Results and Discussion}
\label{sec_discussion}
\subsection{Observational Constraints to $Q_{HII}$}
\label{sez_vincoli_osservativi}
It is known that the intensity of Ly$\alpha$ emission line rapidly decreases
over $z\approx6$~\cite{Dijkstra}. This fact is typically explained with the appearance of neutral hydrogen pockets in the IGM, as shown in some relevant past works available in literature. Here we present a brief summary of some recent observational constraints on $Q_{HII}$.

Mason et al. (2019) found a lower limit of $Q_{HI}=1-Q_{HII} > 0.76$, starting from the study of Ly$\alpha$ emission in a sample of 29 LBGs with photometric redshift within $z = 7.9\pm 0.6$, plus 8 LBGs already observed at $z \approx 8$~\cite{Mason19}. The previous paper by Mason et al. (2018) is based indeed on a cosmological simulation which describes the IGM large scale structures during the EoR. The analysis of damping wings absorption along different line of sight allows them to link its simulated properties to observations on a set of 68 LBGs (collected by Pentericci et al. (2014)~\cite{Pent}), with $-22.75 < M_{UV} < -17.8 $. The resulting HII volume fraction is $0.59_{-0.15}^{+0.11}$~\cite{Mason18}. The same model was than reimplemented in Hoag et al. (2019), with a different LBGs sample, leading to a neutral hydrogen fraction of $Q_{HI} = 0.88^{+0.05}_{-0.10}$, in $z = 7.6 \pm 0.6$, which clearly points to a late-Reionization scenario~\cite{Hoag19}.

In Section \ref{sec_implications_reionization}, we present under the name ``Ly$\alpha$ LF'' a series of publications based on the LAEs luminosity function: Konno et al. (2014), estimated $0.3<Q_{HI}<0.8$ at $z = 7.3$, Konno
et al. (2017) measured $Q_{HI} = 0.3 \pm 0.2$ at $z = 6.6$ on a group of 1266 LAEs, and Zheng et al. (2017) found $0.4 < Q_{HI} < 0.6$ at $z = 6.9$~\cite{Konno14,Konno17,Zheng17}. Mesinger et al. (2015) used a color selected sample of 56 galaxies from Very Large Telescope spectroscopy; connecting large-scale semi-numeric simulations of Reionization with moderate-scale hydrodynamic simulations of the ionized IGM and performing a study on more than 5000 lines of sight, they set an upper limit of $Q_{HII}(z\approx7)\leq0.6$, within a 68$\%$ confidence level~\cite{Mesinger15}. Using the same sample of Konno et al. (2018), Ouchi et al. (2017) obtained the constraint of $Q_{HI} = 0.15 \pm0.15$ at $z = 6.6$, by the comparison between clustering measurements of LAEs and two independent theoretical models~\cite{Ouchi17}. Finally, we report the results by Schenker et al. (2014), which used a set of 451 LBGs to study the correlation between the UV continuum slope of a galaxy and its Ly$\alpha$ emission. Combining a cosmological simulations with Ly$\alpha$ visibility data, they constrained a neutral fraction of $Q_{HI}=0.39^{+0.08}_{-0.09}$, which suggests important evidence that cosmic Reionization ended at $z \approx 6.5$~\cite{Schenker14}.

Following the evolution of $Q_{HII}$ with $z$, we compute the electron scattering optical depth with Equation (\ref{eq_taues}). The results are compared with the CMB measurements performed by Planck and WMAP. In Table \ref{tabella_CMB} we summarize the observational constraints on $\tau_{es}$ and the corresponding instantaneous redshift of Reionization~\cite{Planck15,Planck18,WMAP9}. Discrepancies are due to instrumental systematics and uncertainties on polarized foreground, which are comparable to the statistical error in the WMAP release~\cite{Planck15}. Systematic biases result also from foreground template and masking choice; in particular they involve the treatment of synchrotron and dust emission~\cite{Weiland}.
\begin{specialtable}[H] 
\caption{Electron scattering optical depth and instantaneous redshift of Reionization, measured by WMAP and Planck.}
\label{tabella_CMB}
\setlength{\cellWidtha}{\columnwidth/3-2\tabcolsep+0.0in}
\setlength{\cellWidthb}{\columnwidth/3-2\tabcolsep+0.0in}
\setlength{\cellWidthc}{\columnwidth/3-2\tabcolsep+0.0in}
\scalebox{1}[1]{\begin{tabularx}{\columnwidth}{
>{\PreserveBackslash\centering}m{\cellWidtha}
>{\PreserveBackslash\centering}m{\cellWidthb}
>{\PreserveBackslash\centering}m{\cellWidthc}}
\toprule
\textbf{Name}	& \textbf{\boldmath{$\tau_{es}$}}	& \textbf{\boldmath{$z_{reion}$}} \\
\midrule
WMAP 9	  & $0.088 \pm 0.013$ & $10.5\pm1.1$\\
Planck 2015 & $0.066\pm0.016$ & $8.8^{+1.7}_{-1.4}$\\
Planck 2018     & $0.054\pm0.007$   & $7.7\pm0.7$  \\  
\bottomrule
\end{tabularx}}
\end{specialtable}


\subsection{Initial Condition for Late Reionization Scenario}
\label{sez_condizioni_iniziali}
The evolution of the filling fraction with cosmic time depends also on the initial condition for Equation (\ref{eq_fillingfrac}). In particular, we choose two extreme possibilities, which are motivated both with model available in literature and with an observational point of view.

The first has $Q_{HII}(z=10)=0.2$. This assumption agrees with the $68\%$ credibility interval modelled on the marginalized distribution of the
neutral fraction ($1-Q_{HII}$), from the SFR histories and the Planck constraints on $\tau_{es}$, from Robertson et al. (2015)~\cite{Robertson}.
Similarly, it is coherent with the range of $Q_{HII}$ allowed for the model by Bouwens et al. (2015), where Reionization is complete between $z = 5.9$ and $z = 6.5$~\cite{Bouwens15}.

As a second possibility, we choose $Q_{HII}(z=9)=0.0$, which is preferred by the two hydrogen neutral fraction measurements performed by Mason et al. (2019) and Hoag et al. (2018) and discussed in Section \ref{sez_vincoli_osservativi}~\cite{Hoag19,Mason19}. All the others are intermediate cases.
\subsection{Reionization with Galaxies Only}
\label{sec_reion_gal_only}
\subsubsection{Implications on $f_{esc}$}
\label{sez_implications_fesc}
 
For each of the two initial conditions we compute the number density of ionizing photons $\dot{N}_{ion}$ with different combinations of $\xi_{ion}f_{esc}$, exploring the effect of the parameters degeneracy on the reheating of IGM. 

Particles free-streaming has consequences on galaxy formation, determining a lack of faint-galaxies which alters the UV LF (see Figure \ref{fig_funzioneluminosità}), with a general reduction in the UV luminosity density in models with a high $M_{hm}$. Thus, we obtain a delay in the IGM ionizing process, with respect to CDM.

In Figure \ref{fig_degenerazione}, we show $\log(\xi_{ion}f_{esc})$ in CDM, sterile neutrinos and thermal WDM cosmologies. Due to the great uncertainty on $f_{esc}$, we searched for the $\xi_{ion}f_{esc}$ values that ensure the completion of Reionization at $z=6.7$. We note that $\log(\xi_{ion}f_{esc})$ increases with $M_{hm}$:
a larger escape fraction and/or UV photons production efficiency are needed to complete the Reionization process in WDM scenarios.

\begin{figure}[H]
\includegraphics[width=10.5 cm]{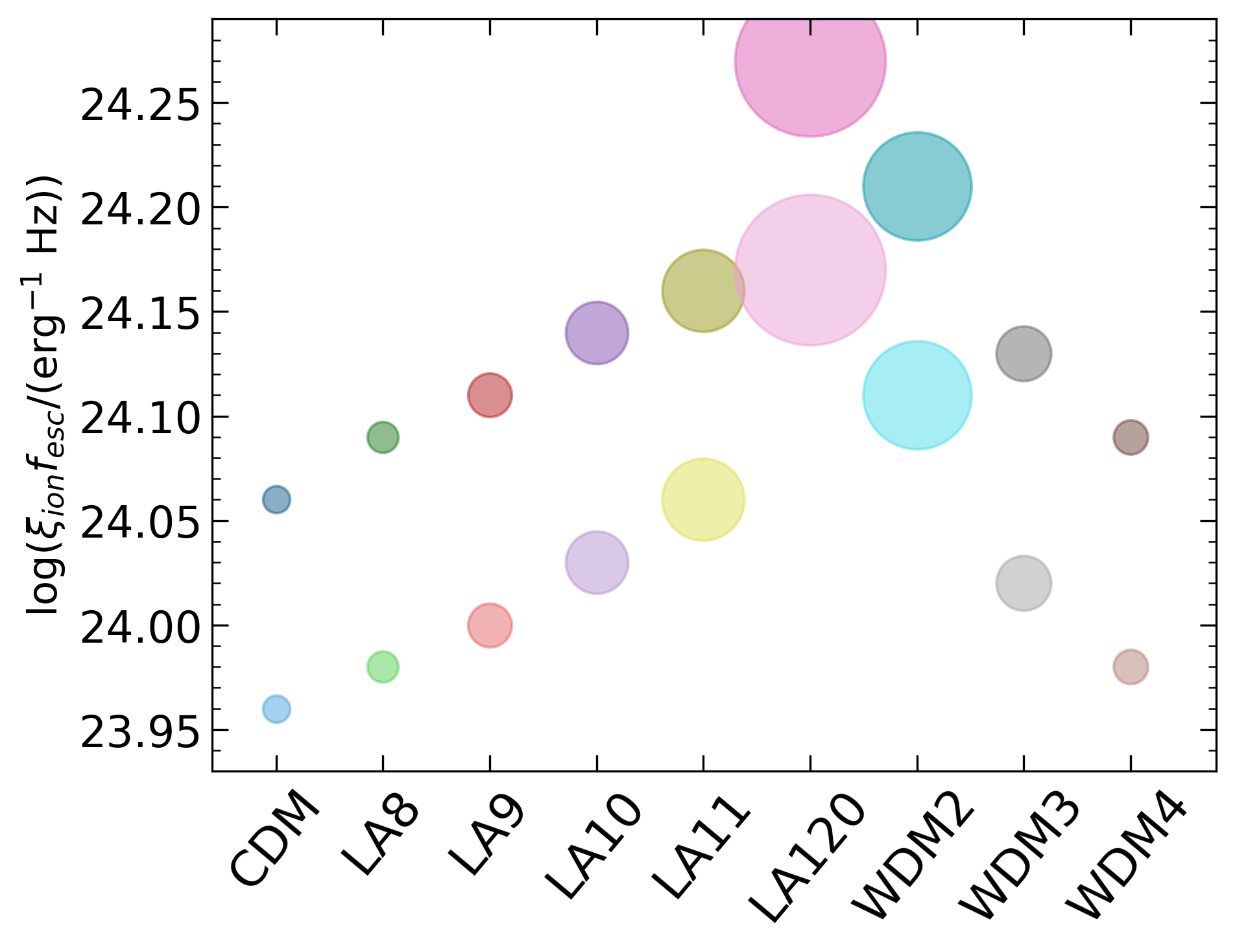}
\caption{The panel shows the values of the product $\xi_{ion}f_{esc}$ required to ionize the IGM at $z=6.7$, for a set of different cosmologies. The dot size increases with $M_{hm}$; lighter colours refer to initial condition $Q_{HII}(z=10)=0.2$, while darker colours are for $Q_{HII}(z=9)=0.0$.}
\label{fig_degenerazione}
\end{figure}

However, the quantity $\xi_{ion}$ is better constrained than $f_{esc}$, so we assume from the literature a fiducial value of $\log(\xi_{ion}/(erg^{-1}Hz))=25.2$~\cite{Finkelstein,Bouwens15,Bouwens16,Yung20}, as expected from a low metallicity single-star population. This value is coherent with the Salpeter IMF assumed in the semi-analytic model~\cite{Robertson,Wilkins} (see Section \ref{subsec_IMF}). We did not investigate the variation of $\xi_{ion}$ with redshift and $M_{UV}$, which we have considered negligible with respect to changes in escape fraction. Similarly, we have neglected the variation with galaxy age. These hypotheses allow us to set an upper limit to $f_{esc}$ for each different WDM particle and boundary condition. In general, models that start from $Q=0$ need a higher $f_{esc}$ value to ionize the IGM within the same $z$ range. For this reason they are more inclusive and result in a weaker constraint to the admitted escape fraction. If $f_{esc}>f_{esc}^{sup}$, Reionization process is completed outside the fiducial redshift interval.

\textls[-15]{In Table \ref{tabella_degenere} we summarize our results for the two initial condition discussed in \mbox{Section~\ref{sez_condizioni_iniziali}}.}

\begin{specialtable}[H] 
\caption{Product $\log(\xi_{ion}f_{esc}/(erg^{-1}Hz)) $ and upper limit of $f_{esc}$, with $\xi_{ion}=10^{25.2} Hz/erg$, for different cosmologies and different initial conditions. We note that these two quantities increase with the half-mode mass $(M_{hm})$. The case [1] refers to $Q_{HII}(z=10)=0.2$. The case [2] refers to $Q_{HII}(z=9)=0.0$.}
\label{tabella_degenere}
\setlength{\cellWidtha}{\columnwidth/6-2\tabcolsep+0.0in}
\setlength{\cellWidthb}{\columnwidth/6-2\tabcolsep+0.0in}
\setlength{\cellWidthc}{\columnwidth/6-2\tabcolsep+0.0in}
\setlength{\cellWidthd}{\columnwidth/6-2\tabcolsep+0.0in}
\setlength{\cellWidthe}{\columnwidth/6-2\tabcolsep+0.0in}
\setlength{\cellWidthf}{\columnwidth/6-2\tabcolsep+0.0in}
\scalebox{1}[1]{\begin{tabularx}{\columnwidth}{
>{\PreserveBackslash\centering}m{\cellWidtha}
>{\PreserveBackslash\centering}m{\cellWidthb}
>{\PreserveBackslash\centering}m{\cellWidthc}
>{\PreserveBackslash\centering}m{\cellWidthd}
>{\PreserveBackslash\centering}m{\cellWidthe}
>{\PreserveBackslash\centering}m{\cellWidthf}}
\toprule
\textbf{Name}	& \textbf{\boldmath{$M_{hm}(M_\odot)$}}	& \textbf{\boldmath{$\log(\xi_{ion}f_{esc})[1]$}} & \textbf{\boldmath{$f_{esc}^{sup}[1]$}} & \textbf{\boldmath{$\log(\xi_{ion}f_{esc})[2]$}} & \textbf{\boldmath{$f_{esc}^{sup}[2]$}}\\
\midrule
CDM		&-			       & 23.96  & 0.06 & 24.06 & 0.07\\
LA8   & $1.3\times10^8$    & 23.98 &   0.06 &  24.09   & 0.08\\
LA9   & $2.6\times10^8$     & 24.00   & 0.06 & 24.11  & 0.08     \\  
LA10  & $5.3\times10^8$     & 24.03 &   0.07 & 24.14  & 0.09  \\
LA11  & $9.2\times10^8$     &  24.06 &  0.07 & 24.16  & 0.09 \\
LA120 & $3.1\times10^9$     & 24.17 &  0.09 & 24.27  & 0.12 \\
WDM 2 & $1.6\times10^9$    & 24.11  &  0.08 & 24.21  & 0.10  \\
WDM 3 & $4.1\times 10^8$    & 24.02  & 0.07 & 24.13  &  0.08 \\
WDM 4 & $1.6\times 10^8$   &23.98  & 0.06  & 24.09 & 0.08 \\

\bottomrule
\end{tabularx}}
\end{specialtable}
\subsubsection{Dependence on the IMF}
\label{subsec_IMF}
The choice of the initial mass function has two main consequences on our study, because it can alter both the UV luminosity emitted by galaxies and the Lyman-continuum photons production efficiency parameter.

\textls[-15]{The UV luminosity is linked to the star formation rate through $SFR=K_{UV}L_{UV}$~\cite{Madau_Dickinson}. For a given SFR, we can correct the UV luminosity by dividing by $K_{UV}(IMF)/K_{UV}(Salpeter)$,} that is 0.63 (Salpeter to Chabrier) or 0.67 (Salpeter to Kroupa); this quantity is nearly constant over a factor of 100 in metallicity~\cite{Madau_Dickinson}. The rise of the UV radiation fields reflects the flattening of the IMF power law towards the lowest stellar masses. 

Some studies predict the value of LyC production efficiency using stellar population synthesis (SPS) codes, which combine evolution and atmosphere models, for a given IMF. In particular, uncertainties on $\xi_{ion}$ are related to treatment of convection, rotation, and evolution of binary stars~\cite{Yung20}. Wilkins et al. (2016) noted that changes in the low-mass end of the Salpeter IMF have a minimal effect on $\xi_{ion}$, in five SPS models~\cite{Wilkins}. Yung et al. (2020) developed a semi-analytic approach with a Chabrier IMF and compared their results with previous studies, including the hydrodynamic simulations presented in Wilkins et al. (2016). They found that $\xi_{ion}$ was nearly unaffected by the change from Chabrier to Salpeter IMF~\cite{Yung20}. Despite the different IMF, also their values agree with our fiducial $\xi_{ion}=10^{25.2} Hz/erg$ assumption. Finally, Eldridge et al. (2017) demonstrated that $\xi_{ion}$ is sensitive to variation in the upper end of the stellar IMF. In particular, they highlighed a lower photon production efficiency in IMF with power law slope extended up to $100$ M$_\odot$ with respect to similar models extended up to $300$ M$_\odot$~\cite{Yung20,Eldridge}. These evaluations lead us to neglect the variations induced on $\xi_{ion}$ by the IMF.
\subsubsection{Dependence on the Clumping Factor}
\label{sect_clumping_factor}

The clumping factor is a quantity related to the density of the IGM, which determines the escape of radiation from an inhomogeneous medium. In particular, $C_{HII}=\frac{\langle \rho^2\rangle}{\langle \rho\rangle^2}$ is altered in alternative cosmological scenarios; changes in its value modify the gas recombination timescale, by enhancing or weakening the recombination process in a clumpy ionized IGM~\cite{Wise,Trombetti}. The paper by Trombetti et al. (2014) analyses the evolution of the clumping factor in a CDM and a WDM framework. Starting from the CDM power spectrum, WDM models are approximated with a $k_{max}$ cut-off, and the evolution of the clumping factor is computed for $k_{max}=$ 200--500--1000~\cite{Trombetti}. Due to the effect of the particle free-streaming, WDM perturbations are completely canceled beyond a characteristic free-streaming wavenumber, so we set $k_{max}\approx k_{fs}$. Free-streaming length is related to the half-model length scale by $\lambda_{hm}\approx 13.93\: \lambda_{fs}^{eff}\approx 13.93\: \epsilon$~\cite{Schneider}. In Table \ref{tab_fs_clump} we show the free-streaming length and the evolution of $C_{HII}$ with $z$. We use the broad limits $k_{max}=200$ for WDM 2 and $k_{max}=1000$ for CDM.
\begin{specialtable}[H] 
\tablesize{\footnotesize}
\caption{\textbf{(a)}: {free-streaming} length for thermal WDM models. \textbf{(b)}: evolution of $C_{HII}$ with $z$ for our fiducial and for the Trombetti model~\cite{Trombetti}.}

\setlength{\cellWidtha}{\columnwidth/7-2\tabcolsep+0.0in}
\setlength{\cellWidthb}{\columnwidth/7-2\tabcolsep+0.4in}
\setlength{\cellWidthc}{\columnwidth/7-2\tabcolsep+0.7in}
\setlength{\cellWidthd}{\columnwidth/7-2\tabcolsep-0.3in}
\setlength{\cellWidthe}{\columnwidth/7-2\tabcolsep-0.3in}
\setlength{\cellWidthf}{\columnwidth/7-2\tabcolsep-0.3in}
\setlength{\cellWidthg}{\columnwidth/7-2\tabcolsep-0.2in}
\scalebox{1}[1]{\begin{tabularx}{\columnwidth}{
>{\PreserveBackslash\centering}m{\cellWidtha}
>{\PreserveBackslash\centering}m{\cellWidthb}
>{\PreserveBackslash\centering}m{\cellWidthc}
>{\PreserveBackslash\centering}m{\cellWidthd}
>{\PreserveBackslash\centering}m{\cellWidthe}
>{\PreserveBackslash\centering}m{\cellWidthf}
>{\PreserveBackslash\centering}m{\cellWidthg}}
\toprule

 & \textbf{(a)} & & & \textbf{(b)} & \\
\midrule
\textbf{Name}	& \boldmath{$\lambda_{fs}^{eff}$} \textbf{(Mpc h\boldmath{$^{-1}$})} & \textbf{Model} & \boldmath{$z=7$}&\boldmath{$z=8$}&\boldmath{$z=9$}&\boldmath{$z=10$}\\
\midrule
WDM 2	  & $276$ & $C_{HII}=1+43z^{-1.71}$	  & $2.54$&$2.23$&$2.00$&$1.84$\\
WDM 3 & $433$ & $C_{HII}^{CDM}$ & $3.04$&$2.59$&$2.29$&$2.06$ \\
WDM 4     & $596$ & $C_{HII}^{WDM2}  $   & $2.39$&$2.06$&$1.85$&$1.71$\\  
\bottomrule
\end{tabularx}}
\label{tab_fs_clump}
\end{specialtable}
Figure \ref{fig_clump} shows the evolution of the filling fraction with different $C_{HII}$ models. A fixed $C_{HII}=3$ value implies a reduction in $t_{rec}$ and delay in Reionization process. Our fiducial model (Equation (\ref{eq_clump})) falls between $C_{HII}^{CDM}$ and $C_{HII}^{WDM2}$ for every considered redshift. This yields to an early end of EoR in CDM case and in a further delay in WDM 2 case. The overall effect on the Reionization history is, however, perturbative. 

\begin{figure}[H]
\includegraphics[width=10 cm]{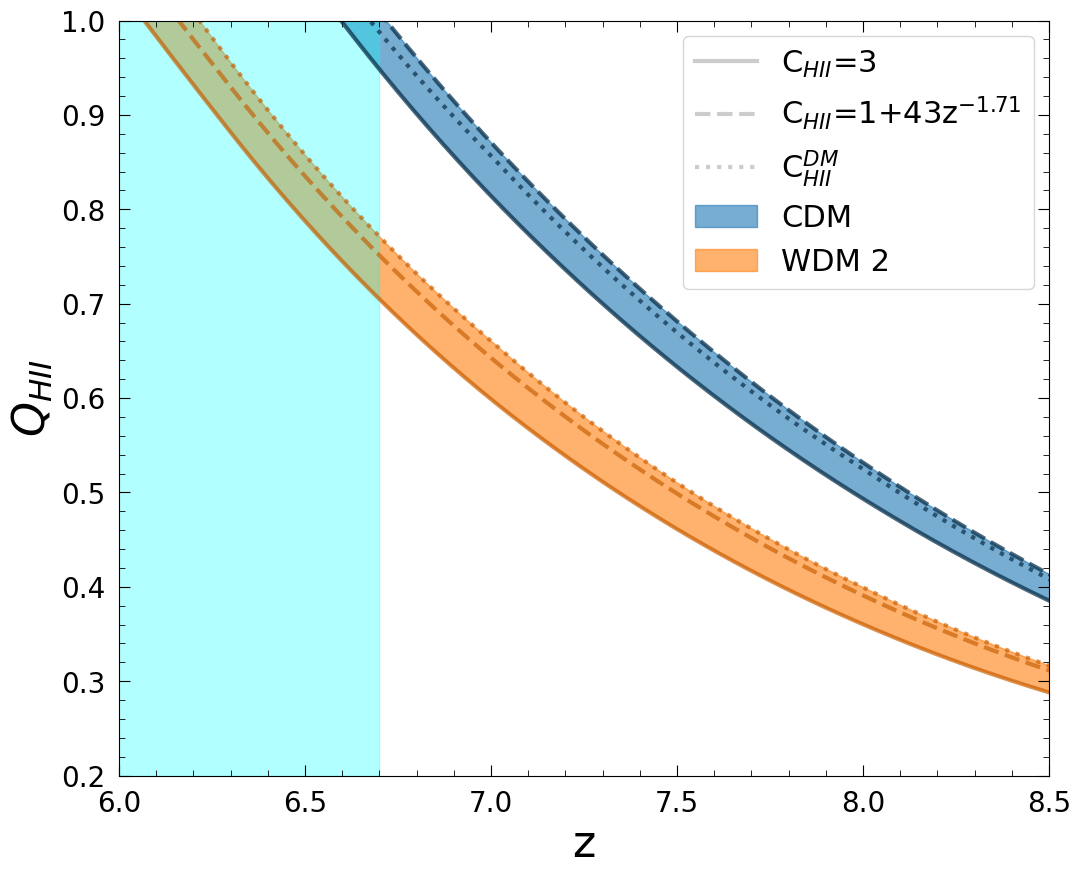}
\caption{Evolution of the filling fraction with different clumping factor models, for two representative cosmologies: the continuous line refers to fixed $C_{HII}=3$, while the dotted and the dashed lines indicate respectively models dependent on redshift and cosmology (Trombetti et al. (2014)~\cite{Trombetti}), and on redshift only.}
\label{fig_clump}
\end{figure}

\subsubsection{Implications on the Reionization History}
\label{sec_implications_reionization}
In our reconstruction of EoR, we found that the bulk of Reionization happens between $6<z<8$, with significant differences which derive from initial condition and $f_{esc}$ model.

In Figure \ref{fig_filling_tau_fixed} we show the evolution of the filling fraction in sterile neutrino and thermal WDM scenarios, with $\log(\xi_{ion}/(erg^{-1}Hz))=25.2$ and fixed escape fraction. We note that the Reionization pattern changes with respect to the initial conditions. Models with $Q_{HII}(z=9)=0.0$, which have $f_{esc}=0.06$, show a better agreement with the data, not only with respect to Hoag19 and Mason19 measurements, but also with the other observational points. Differences between cosmologies are tiny, but increase with the age of universe. However, they do not allow us to discriminate between CDM and WDM scenarios with the actual experimental uncertainties. We also point out that when $Q_{HII}(z=10)=0.2$, the escape fraction is lower ($f_{esc}=0.05$), because the IGM is already partial ionized when we start to solve Equation (\ref{eq_fillingfrac}).

\begin{figure}[H]
\includegraphics[width=12 cm]{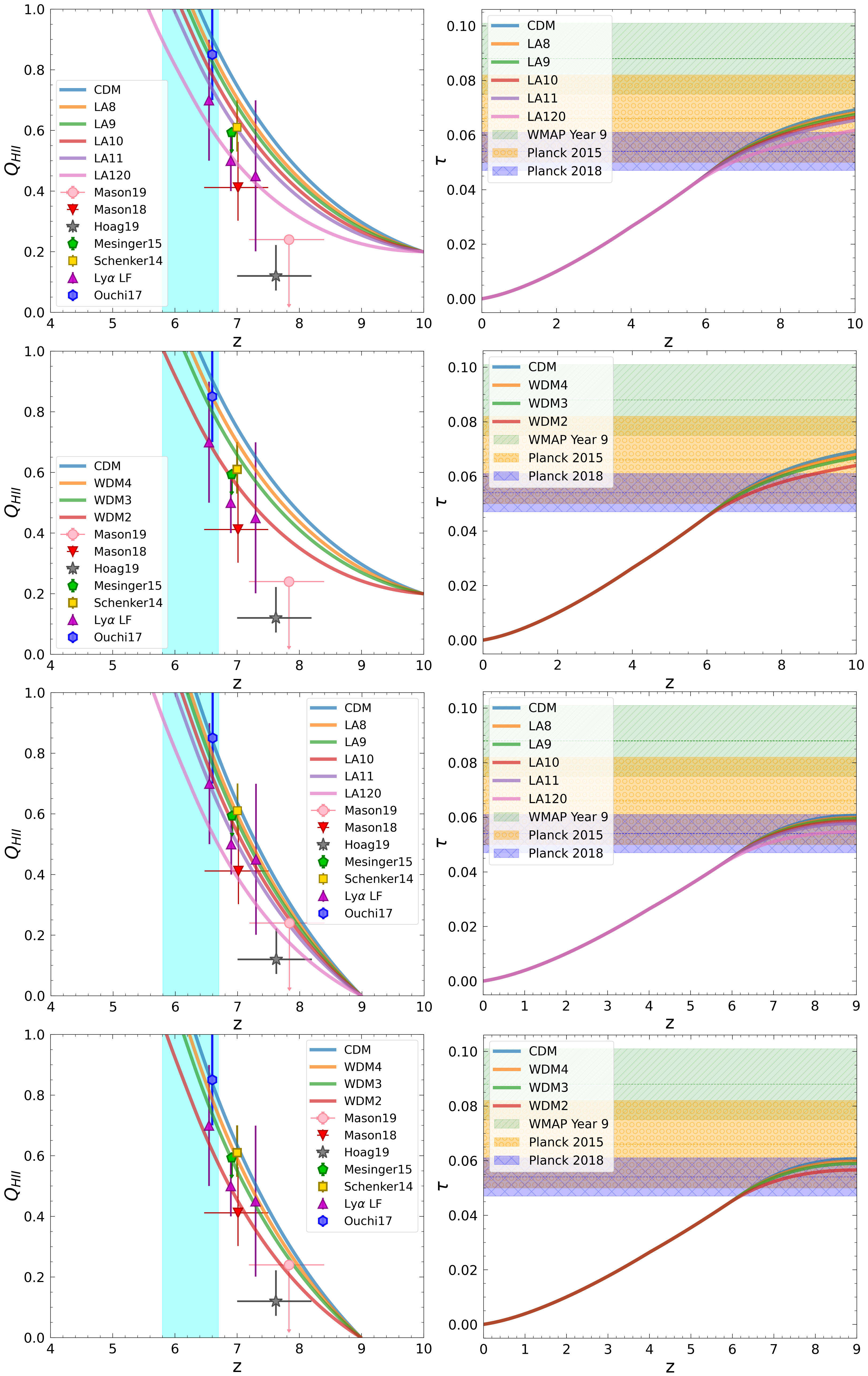}
\caption{(\textbf{Left}): {evolution of the filling} fraction $Q_{HII}$, for sterile neutrino and thermal WDM models, with $\log(\xi_{ion}/(erg^{-1}Hz))=25.2$. The two upper panels have initial condition $Q_{HII}(z=10)=0.2$ and $f_{esc}=0.05$. The two lower panels are plotted with $Q_{HII}(z=9)=0.0$ and $f_{esc}=0.06$. The cyan shaded region indicates our fiducial late-Reionization redshift interval, $5.8<z<6.7$. The upward triangle labelled Ly$\alpha$ LF includes results by Konno et al. (2014), Konno et al. (2017) and Zheng et al. (2017)~\cite{Konno14,Konno17,Zheng17}. (\textbf{Right}): electron scattering optical depth for different models, compared with measurements from Planck and WMAP~\cite{Planck15,Planck18,WMAP9}.} \label{fig_filling_tau_fixed}
\end{figure}  

In the right panels, the analysis of electron scattering optical depth shows a $1\sigma$ agreement with the Planck 2015 result. In particular, if $Q_{HII}(z=9)=0$, $\tau_{es}$ saturates in most models within the range of Planck 2018 measurements. Conversely, we should note that cosmological scenarios with $Q_{HII}(z=10)=0.2$ cannot reach their definitive optical depth value, which depends on how the $Q_{HII}(z)$ decreases at $z>10$. This latter aspect highlights the main limit of our semi-analytic model. In fact, as shown in Section \ref{subsec_funzione_lum}, it can produce the UV LF only at $z\lesssim8$; the evolution of the filling fraction is basically obtained with a high redshift extrapolation of the ionizing photons production rate, which becomes less reliable towards previous redshifts.

In recent years, several studies have introduced a dependence between $f_{esc}$, halo-mass and redshift. In most of them, the escape fraction progressively decreases in systems at low $z$ and with a larger $M_{halo}$. From a physical point of view, this fact is justified as a consequence of the lower gaseous column density in less massive halos, which are also more abundant in the young universe. 

As comparison, we have tested some relevant models from the literature. The $f_{esc}(M_{halo})$ model computed in Yajima et al. (2011) combines the results of hydrodynamic simulations with radiative transport calculations of stellar radiation and evaluates the effects of dust extinction, for mass halos in the range $10^9$--$10^{11}$ $M_\odot$~\cite{Yajima}. They highlighted that $f_{esc}$ shows a noticeable decrease as $M_{halo}$ increases, in every redshift bin between 3 and 6; however, it is not significantly altered by $z$~\cite{Yajima}. So, we freeze the $f_{esc}(z=6)$ values (\mbox{Table~\ref{tab:Yajima}}) and we use them in the rest of our analysis. 
While this, and similar models, represent the state-of-the-art in the modelling of $f_{esc}$, it is yet unknown whether their assumptions hold up at very high redshifts. However, lacking observational and theoretical evidence in support of different assumptions, the adoption of this halo-mass-dependent escape fraction is surely instructive because it shows the macroscopic consequences that we obtain by increasing the relative contribution of faint galaxies and, at the same time, by reducing the relative contribution of the brightest ones. 
\begin{specialtable}[H] 
\caption{Table from Yajima et al. (2011)~\cite{Yajima} with dependence between escape fraction and halo-mass.}
\setlength{\cellWidtha}{\columnwidth/2-2\tabcolsep+0.0in}
\setlength{\cellWidthb}{\columnwidth/2-2\tabcolsep+0.0in}
\scalebox{1}[1]{\begin{tabularx}{\columnwidth}{
>{\PreserveBackslash\centering}m{\cellWidtha}
>{\PreserveBackslash\centering}m{\cellWidthb}}

\toprule
\textbf{\boldmath{$\log M/M_\odot$}}	& \textbf{\boldmath{$f_{esc}(z=6)$}}	\\
\midrule
     9     & 0.325 \\
    9.5   & 0.212 \\
    10    & 0.115 \\
    10.5  & 0.132 \\
    11    & 0.031 \\
\bottomrule

\end{tabularx}}
\label{tab:Yajima}
\end{specialtable}
\vspace{-8pt}
In our semi-analytic model, the bulk of ionizing photons is produced within halos with $\log(M/M_\odot)<10.5$. This fact is strongly enhanced by the introduction of Yajima escape fraction, which almost suppresses the contribution of $\log(M/M_\odot)>10.5$ halos. The effect of galaxy-galaxy merging is neutralized by the reduction of $f_{esc}$ in high-mass systems. So, the contribution of bright galaxies drops and they cannot imprint the characteristic acceleration that we see when $f_{esc}$ is fixed. Furthermore, the model emphasizes the differences between cosmologies, which are related to the number density of the faint sources in each scenario. However, the $f_{esc}$ value remains larger than $10\%$ for most of the mass bins; a higher escape fraction requires at the same time to reduce $\xi_{ion}$ to $10^{24.9} Hz/erg$, in order to complete the Reionization process within our fiducial $\Delta z$. 

Differently, Puchwein et al. (2019) adopted the functional form: $f_{esc}(z) = \mathrm{min} [6.9 \times 10^{-5}(1+z)^{3.97} , 0.18]$, which for the authors is preferred by direct measurements and estimates of the HI photoionization rate~\cite{Puchwein}. This prescription gives $16\%<f_{esc}<18\%$ in the interval $6<z<10$ and determines an early completion of the process despite the $\log(\xi_{ion}/(erg^{-1}Hz))=24.9$ value. Thus, as reported in Figure \ref{fig_filling_tau_yaj}, the joint use of our semi-analytic model, with both the variable escape fractions presented by Yajima et al. (2011) and Puchwein et al. (2019), is reasonably disfavoured. 
\begin{figure}[H]	
\includegraphics[width=10.5 cm]{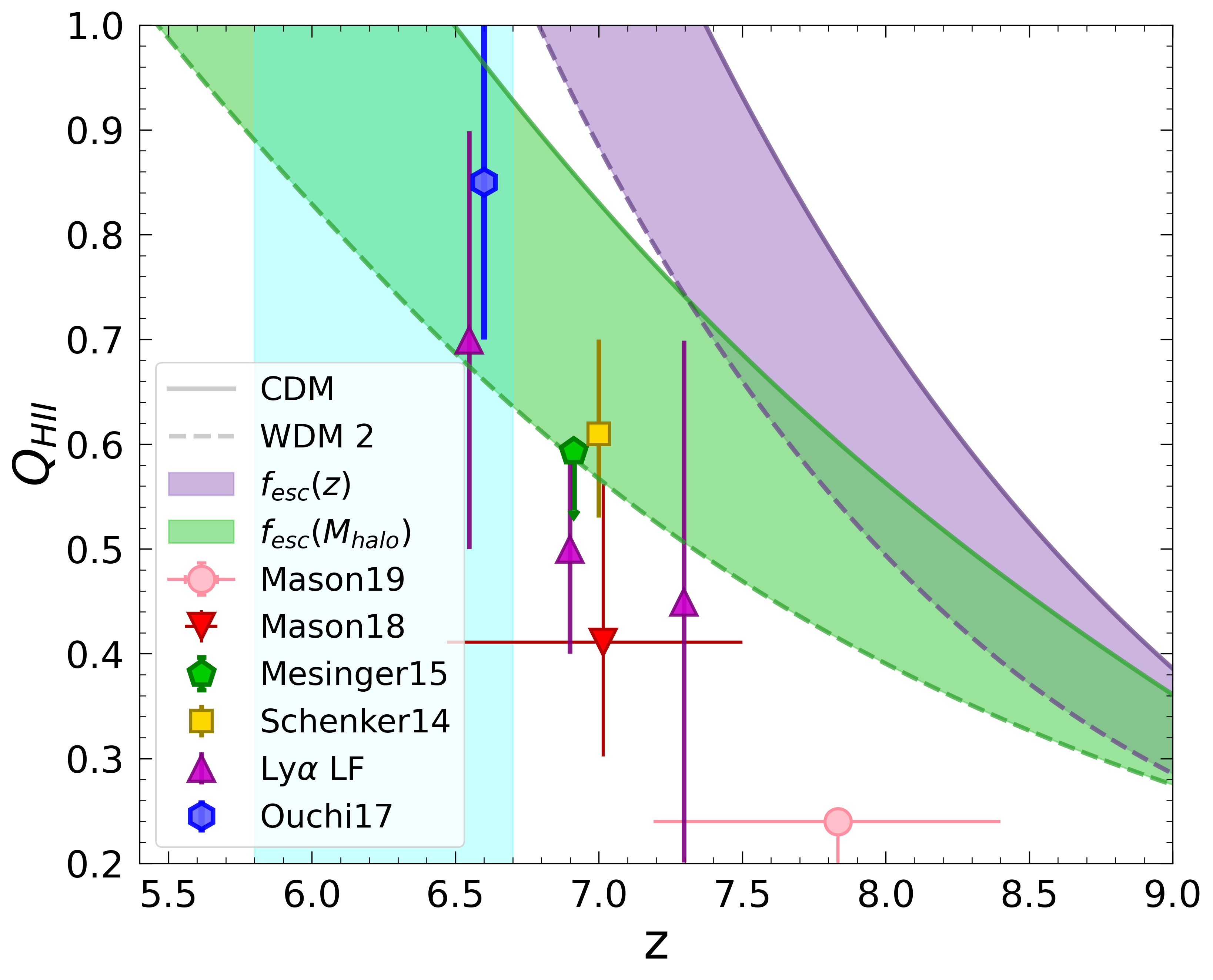}
\caption{Evolution of the filling fraction $Q_{HII}$, for CDM and WDM 2 models, with $f_{esc}(M_{halo})$ as computed by Yajima et al. (2011)~\cite{Yajima} and $f_{esc}(z)$ by Puchwein et al. (2019)~\cite{Puchwein}, with $\log(\xi_{ion}/(erg^{-1}Hz))=24.9$. The cyan shaded region indicates our fiducial late-Reionization redshift interval, $5.8<z<6.7$. The upward triangle labelled Ly$\alpha$ LF includes results by Konno et al. (2014), Konno et al. (2017) and Zheng et al. (2017)~\cite{Konno14,Konno17,Zheng17}.}
\label{fig_filling_tau_yaj}
\end{figure} 
\subsection{Comparison with Previous Works on Reionization}
\label{sez_confronto_finale}
In this work, we have used different assumptions about the escape fraction value. In the hypothesis that galaxies are the only ionizing source during the EoR, we find an upper limit to $f_{esc}$ for each considered cosmology. Depending on initial condition, it can varies between 0.06 and 0.12. However, $f_{esc}$ is a very low-constrained parameter in literature; it depends from a huge number of physical properties and there are many studies that offer their proper estimation (see Dayal P. and Ferrara A. (2018) for a review~\cite{Dayal_RevieW}).
It is interesting to discuss this result in the context of previous papers about the \mbox{EoR. Carucci et al. (2018)} concluded that different cosmologies have similar Reionization histories, and current probes are insensitive to the specifics of DM scenarios. In particular, they focused on the effect of degenerate model parameters $M_{UV}^{lim}$ and $f_{esc}$: for each DM model, assuming $M_{lim}^{UV}\geq -13$, the escape fraction lies between $0.07\lesssim
f_{esc}\lesssim 0.15$ at $2\sigma$~\cite{Carucci}. Dayal et al. (2020) found a final Reionization redshift between $5<z<6.5$, analyzing both the role of galaxies and AGNs~\cite{Dayal20}. Finkelstein et al. (2019) completed the Reionization at $z\approx6$, using a halo-mass dependent low escape fraction~\cite{Finkelstein}. In these two galaxy-driven works, AGNs become dominant at $z\lesssim5$. Moreover, Bouwens et al. (2015) concluded that $Q_{HII}$ reaches the unity between $5.9<z<6.5$~\cite{Bouwens15}, while Robertson et al. (2015), using $f_{esc}=0.2$, highlighted an evolution of the filling fraction from $0.2$ to $0.9$ during $6<z<9$~\cite{Robertson}, with $\tau_{esc}$ within the Planck 2015 $1\sigma$ range result. Finally, we remind the work by Dayal et al. (2017), which suggests that, in CDM case, the bulk of the ionizing photons production is related to low-mass systems with $M_{halo}<10^9 M_\odot$ and $-15<M_{UV}<-10$, while in $1.5$ and $3$ keV thermal WDM it is shifted towards $M_{halo}>10^9 M_\odot$ and $-17<M_{UV}<-13$~\cite{Dayal17}. This result is slightly different from our galaxies-only Reionization model, which instead admits a significant contribution from $M_{halo}>10^{10.5} M_\odot$ and $M_{UV}\gtrsim-20$ sources. 
\section{Summary and Conclusions}
\label{sec_conclusioni}
In this paper we have studied how the suppression of the CDM power spectrum due to particles free streaming can have macroscopic consequences, via a late galaxy formation, on the Reionization of IGM. We have used the semi-analytic model described by Menci et al. (2018)~\cite{Menci18}, to produce the UV LF in a $\Lambda$CDM framework. We have tested some $\Lambda$WDM cosmologies, in which the contribution of the faint galaxies is suppressed: in particular, we have focused on five sterile neutrino models (LA8-LA9-LA10-LA11-LA120, presented in Lovell et al. (2020)~\cite{Lovell}), with the same particle mass, but with different mixing angle, and three thermal WDM models (WDM2-WDM3 and WDM4), with, respectively, \mbox{$m_X=$ 2--3--4 keV}. In both cases, we have found that a higher $M_{hm}$ leads to a general delay in the Reionization process.

We have explored the hypothesis that the hydrogen reheating is driven by galactic photons only. In this context, the relative intrinsic emission of faint and bright sources and the differences in Reionization history are related to the specific escape fraction assumption. In particular, if $f_{esc}$ is fixed:
\begin{itemize}
    \item in CDM cosmology, merging phenomena between halos increase the relative contribution of systems with $M_{halo}>10^{10.5} M_\odot$, from $\approx$$22\%$ at $z=8$ up to $\approx$$48\%$ at $z=6$. At the same time, in WDM scenarios, the role of low mass systems is reduced, depending on $M_{hm}$;
    \item in CDM cosmology, merging between galaxies determines the rise of the intrinsic $M_{UV}<-20$ systems relative contribution to the ionizing photons budget, from $\approx$$30\%$ to $\approx$$45\%$ between $6.3<z<8$. However, it remains subdominant during the EoR, because faint galaxies with $M_{UV}>-20$ emit the bulk of ionizing photons. On the other hand, in WDM case, the particles free-streaming yields to a shift towards brighter sources and $r_{phot}(M_{UV}<-20)$ undergoes a further 1--10$\%$ growth, depending on cosmology;
    \item in WDM cosmologies a higher $f_{esc}\xi_{ion}$ is required, in order to complete the Reionization process at the same redshift. Using $\log(\xi_{ion}/(erg^{-1}Hz))=25.2$ we set an upper limit $f_{esc}^{sup}$ for the Reionization at $z=6.7$, which goes, depending on $M_{hm}$ and on the high-z $Q_{HII}$ value, from $0.06$ in the $\Lambda$CDM model to $0.12$ in the LA120. So, in general, WDM scenarios yield to an overall reduction of $\dot{N}_{ion}$ with respect to CDM.
\end{itemize}

The Reionization history depends also on the choice of the initial ionized hydrogen filling fraction: models with $Q_{HII}(z=9)=0$ better agree with observational data, while models with $Q_{HII}(z=10)=0.2$ do not match the neutral fraction measurements performed by Mason et al. (2019) and Hoag et al. (2019)~\cite{Hoag19,Mason19}. We have also evaluated the role of the clumping factor in the reheating of IGM, finding that $C_{HII}$, in WDM cosmologies, is just tiny altered with respect to our fiducial redshift scaling relation.

Finally, we have tested some relevant variable escape fraction available in \mbox{the literature}:  
\begin{itemize}
    \item in the $f_{esc}(M_{halo})$ model by Yajima et al. (2011)~\cite{Yajima}, the impact of systems with $M_{halo}>10^{10.5}M_\odot$ and $M_{UV}<-20$ is suppressed with respect to fixed $f_{esc}$ case, while faint galaxies play a more important role. However, the latter are affected by WDM free streaming and this results in delay in the Reionization process;
    \item the $f_{esc}(z)$ model by Puchwein et al. (2019) does not alter the relative contribution of bright and faint systems, but it produces too many ionizing photons at high redshift;
    \item both models, when implemented in our semi-analytic framework, are reasonably disfavoured by the low $\log(\xi_{ion}/(erg^{-1}Hz))\approx24.9$ value.
\end{itemize}

To conclude, we note that the most important limits to our analysis are related to observational uncertainties. In the future, we expect significant advances in the field of extra-galactic astrophysics and observational cosmology. For example, the role of AGNs during the EoR will be better understood with Euclid, by looking at the Quasar Luminosity Function in the early universe. Also the detection of QSOs damping wing can be used to measure the cosmic neutral hydrogen fraction~\cite{Euclid}. 
Another research area is currently investigating the 21-cm signal produced by the ground-state hyperfine transition of atomic hydrogen. Recent interferometric  observations (LOFAR, PAPER) have identified an upper limit to the 21-cm emission. In the future the high-z tomography performed by SKA (Square Kilometer Array) will be able to reconstruct the distribution and the topology of the neutral hydrogen during the EoR.

Future data releases will offer a better estimation of electron scattering optical depth. In particular, CLASS (Cosmic Large Angular Scale Survey) and LiteBIRD (Lite satellite for the study of  B-mode polarization and Inflation from cosmic background Radiation Detection) are projected to recover $\tau_{es}$ to nearly cosmic variance uncertainties. Although the electron scattering optical depth and the neutral hydrogen fraction provide important constraints on Reionization redshift, further spectroscopic measurements are needed in order to characterize high-z galaxies $f_{esc}$ and $\xi_{ion}$. Moreover, a better estimation of the $z>6$ UV LF will allow us to search for a possible turn-over in its faint-end, and to set constraints on WDM cosmological scenarios. Both goals can be achieved by exploring the deep parts of the universe
that will be accessible to the James Webb Space Telescope.
\vspace{6pt} 



\authorcontributions{Conceptualization, M.R., N.M., M.C.; methodology, M.R., N.M. and M.C.; software, N.M.; validation, N.M., M.C.; data curation, M.R.; writing---original draft preparation, M.R.; writing---review and editing, M.R., N.M. and M.C.; visualization, M.R.; supervision, N.M., M.C.; project administration, N.M., M.C. All authors have read and agreed to the published version of the manuscript.}

\funding{This research received no external funding.}

\institutionalreview{{Not applicable.}
}

\informedconsent{{Not applicable.}

}

\dataavailability{{Not applicable.}
} 

\acknowledgments{We thank N. Sanchez for  inviting us to write the present paper, L. Pentericci for supervision during the work and M. Parente for the aesthetic advice.}

\conflictsofinterest{The authors declare no conflict of interest.} 


\end{paracol}
\reftitle{References}

%


\end{document}